\documentclass[twocolumn,aps,prc,showpacs,superscriptaddress,preprintnumbers,floatfix,nofootinbib]{revtex4}
\usepackage{epsfig,graphics}
\usepackage{graphicx}
\usepackage{dcolumn}
\usepackage{bm}
\usepackage{amsmath}
\usepackage[usenames]{color}
\usepackage{ulem} 

\begin{document}

\title{Initial fluctuation effects on harmonic flows in high-energy heavy-ion collisions }

\author{ L. X. Han} \affiliation{Shanghai Institute of
Applied Physics, Chinese Academy of Sciences, Shanghai 201800,
China}
 \affiliation{Graduate
School of the Chinese Academy of Sciences, Beijing 100080, China}
\author{ G. L. Ma}
\affiliation{Shanghai Institute of Applied Physics, Chinese
Academy of Sciences, Shanghai 201800, China}
\author{ Y. G. Ma}
\thanks{Author to whom all correspondence should be addressed. Email: ygma@sinap.ac.cn}
\affiliation{Shanghai Institute of Applied Physics, Chinese
Academy of Sciences, Shanghai 201800, China}

\author{ X. Z. Cai}
\affiliation{Shanghai Institute of Applied Physics, Chinese
Academy of Sciences, Shanghai 201800, China}
\author{ J. H. Chen}
\affiliation{Shanghai Institute of Applied Physics, Chinese
Academy of Sciences, Shanghai 201800, China}

\author{ S. Zhang}
\affiliation{Shanghai Institute of Applied Physics, Chinese
Academy of Sciences, Shanghai 201800, China}
\author{ C. Zhong} \affiliation{Shanghai Institute of Applied Physics, Chinese
Academy of Sciences, Shanghai 201800, China}
\date{ \today}

\begin{abstract}
Within the framework of a multi-phase transport model, harmonic
flows $v_n$ (n = 2, 3 and 4) are investigated for Au + Au
collisions at $\sqrt{s_{NN}}$ = 200 GeV and Pb + Pb collisions at
$\sqrt{s_{NN}}$ = 2.76 TeV. The event-by-event geometry
fluctuations significantly contribute to harmonic flows.
Triangular flow ($v_3$) originates from initial triangularity
($\varepsilon_3$) and is developed by partonic interactions. The
conversion efficiency ($v_n/\varepsilon_n$) decreases with
harmonic order and increases with partonic interaction cross
section. A mass ordering in the low $p_T$ region  and number of
constitute quark scaling  in the middle $p_T$ region seem to work
roughly for $n$-th harmonic flows at both energies.
All features of harmonic flows show similar qualitative behaviors
at RHIC and LHC energies, which implies that the formed partonic
matters are similar at the two energies.

\end{abstract}


\maketitle

\section{Introduction}

Results from the Brookhaven Relativistic Heavy-Ion Collider (RHIC)
indicate that a strongly-interacting partonic matter has been
created in relativistic nucleus-nucleus
collisions~\cite{RHIC_white_paper}. A powerful probe exposing
the characteristics of new matter, elliptic flow, has been
measured via the second Fourier coefficient ($v_2$) in the
azimuthal distribution of final particles. It is translated from
an early stage coordinate space asymmetry, which can reflect how the hot matter evolves hydrodynamically~\cite{RHIC_white_paper,Voloshin1, Jinhui4}. The $v_2$
data show remarkable hydrodynamical behaviors, which implies the
formed matter is thermalized in a very short time and expands
collectively as a perfect-like liquid with a very small shear
viscosity over entropy density ratio $({\eta}/{s}$)
\cite{eta_s_L,eta_s_R,eta_s_H,eta_s_E}. Elliptic flow ($v_{2}$)
has been studied widely as functions of centrality, transverse
momentum ($p_T$) and pseudorapidity ($\eta$) etc. A mass-ordering at low $p_T$ and a Number of Constituent Quark (NCQ) scaling at intermediate
$p_T$ for $v_2$ have been observed, which suggests that a thermalized partonic
matter is formed and a collective motion is developed prior to
hadronization~\cite{strangeBaryonmassorder5,
STARMassOrder6,PHENIXNCQ7,PHENIXphiNCQ8,tianjianNCQ9}. On the
other hand, a geometry (participant eccentricity) scaling was
observed for $v_2$ fluctuations, which implies not only participant eccentricity is
responsible for elliptic flow, but also the event-by-event initial
state geometry fluctuations contribute to harmonic flow~\cite{Paul,partie2v210,partie2v2fluc11}.

It has been recently found that the triangular flow ($v_3$) is not
zero in the azimuthal distribution of final particles. In fact,
because of the non-smooth profile, coming from the event-by-event
fluctuations of participant nucleons, it shows a triangular
initial geometry shape can be transferred into momentum space by
hydrodynamical evolution. In recent studies, it has been
demonstrated that triangular flow significantly contributes on the
near-side ridge and away-side double bumps in two-particle
azimuthal correlations~\cite{triangularflow12,kotriangularflow13}.
As a new probe, triangular flow is believed to provide more
information about the formed hot and dense matter. It has been
studied as functions of centrality, transverse momentum,
pseudorapidity ($\eta$), as well as the relations with the initial
triangularity ($\varepsilon_3$) and shear viscosity over entropy
density ratio
~\cite{triangularflow12,kotriangularflow13,ampthotspots14,Petersenv3,Qinv3}.
However, the dependence of triangular flow on the elastic two-body
partonic scattering cross section is absent. In addition, a
possible NCQ-scaling, which has been found held by the elliptic
flow~\cite{v4NCQ160}, have not been studied in details for other
$v_n$ ($n$=3,4...) when the initial fluctuations are taken into
account .

This work presents  the initial deformation scaling of elliptic
($v_2$), triangular ($v_3$) and quadrangular flows ($v_4$) for
different cross sections within the framework of the AMPT
model~\cite{AMPT2160,AMPT216}. The mass ordering at low $p_T$ and
constituent quark number scaling at higher $p_T$ for the $v_n$ are
investigated after considering the event-by-event initial state
geometry fluctuations at RHIC and LHC energies. Meanwhile, a
special care is discussed for $s$-quark and $\phi$ meson for
$v_n$-scaling.

The paper is organized in the following way. A brief description
of the AMPT model is introduced in Sec. II. The results and discussions are
presented in Sec. III. Finally, a summary is given in Sec. IV.

\section{Brief description of AMPT model}

A multi-phase transport (AMPT) model consists of four main
components: the initial condition, partonic interactions,
conversion from partonic to hadronic matter, and hadronic
interactions. The initial condition, which includes the spatial
and momentum distributions of minijet partons and soft string
excitations, is obtained from the Heavy Ion Jet Interaction
Generator (HIJING) model. Scatterings among partons are modeled by
Zhang's Parton Cascade (ZPC) model, which  includes only two-body
scatterings with cross sections obtained from the pQCD
calculations with screening mass. In the default version of AMPT
model, partons only include minijet partons, and recombine with
their parent strings when they stop interactions, then the
resulting strings are converted to hadrons by using the Lund
string fragmentation mechanism. While in the version with the
string melting mechanism, partons include minijet partons and
partons from melted strings. And a quark coalescence model is used
to combine partons into hadrons. The dynamics of the subsequent
hadronic matter is then described by a relativistic transport
(ART) model. Details of the AMPT model can be found in a
review~\cite{AMPT216}. Previous AMPT calculations have found that
elliptic flow can be built by strong parton
cascade~\cite{AMPT216,SAMPT17,AMPT_v2_e250,AMPTLHC} and jet losses
energy into partonic medium to excite an away-side double-peak
structure~\cite{ampthotspots14, amptdihadron19,amptgammajet20}.
Therefore, partonic effect can not be neglected and the string
melting AMPT version is much more appropriate than the default
version when the energy density is much higher than the predicted
critical density . In this work, we use the version of AMPT model
with the string melting mechanism to simulate Au+Au collisions at
$\sqrt{s_{NN}}$ = 200 GeV as well as Pb + Pb collisions at
$\sqrt{s_{NN}}$ = 2.76 TeV.  Since collective flow has been built
up after the expansion of partonic stage, we neglect the final
hadronic rescattering effects on harmonic flows in this work.

\section{Results and Discussions}

\subsection{Brief definition of $v_n$ with initial fluctuations}

We know harmonic flows are defined as the $n$-th Fourier coefficient
$v_n$ of the particle distribution with respect to the reaction plane.
However, after considering event-by-event fluctuations in the initial density distribution~\cite{triangularflow12}, the particle distribution should be written as
      \begin{equation}
      \frac{dN}{d\phi}
      \propto1+2\sum_{n=1}^{\infty}v_ncos[n(\phi-\psi_n)],
      \label{eq1}
      \end{equation}
where $\phi$ is the momentum azimuthal angle of each hadron.
$\psi_{n}$ is the $n$-th event plane which varies due to
event-by-event fluctuations and can be calculated by
\begin{equation}
\psi _n^{r}  = \frac{1}{n}\left[ \arctan\frac{\left\langle {r^2 \sin
(n\varphi)} \right\rangle}{\left\langle {r^2 \cos (n\varphi)}
\right\rangle} + \pi \right],
 \label{eq3}
\end{equation}
where $r$ and $\varphi$ are the coordinate position and azimuthal angle
of each parton and the average $\langle \cdots\rangle$ is density weighted in the initial state, and and the superscript $r$ denotes initial coordinate space. The
$n$-th order eccentricity $\epsilon_{n}$ for initial geometric distribution is defined as

\begin{equation}
\varepsilon _n  = \frac{{\sqrt {\left\langle {r^2 \cos (n\varphi
)} \right\rangle ^2 + \left\langle {r^2 \sin (n\varphi)}
\right\rangle ^2 } }}{{\left\langle {r^2 } \right\rangle }}.
\label{eq2}
\end{equation}

There is some arbitrariness in the definition of $\psi_n(r)$ and
$\varepsilon_n$ \cite{Oll}, because one could, for instance,
replace $r^2$ with $r^n$ in Eq. (2) and (3) \cite{Li}.  With this
replacement, however, $v_3$ and $v_4$  only change little in our
calculations (less than 3\% for $v_3$ and 12\% for $v_4$,
respectively, for 0-80\% centrality). In the following
calculations, we will use Eq.(\ref{eq3}) and Eq.(\ref{eq2}) to
decide $\psi_n$ and $\varepsilon _n$.

After $\psi _n$ is determined, the $n$-th harmonic flow $v_n$ can be obtained by
      \begin{equation}
      v_n^{r} = \left\langle cos[n(\phi-\psi_n^{r})] \right\rangle.
      \label{eq4}
      \end{equation}


In alternative way, $\psi_n$ and $v_n$ can be also calculated in momentum space as,

\begin{equation}
\psi_n ^{p}  = \frac{1}{n}\left[ \arctan\frac{\left\langle {p_{T}
\sin (n\phi)} \right\rangle}{\left\langle {p_{T} \cos (n\phi)}
\right\rangle} \right],
 \label{eq5}
\end{equation}
and
\begin{equation}
v_n^{p} = \left\langle cos[n(\phi-\psi_n^{p})] \right\rangle,
\label{eq6}
\end{equation}
where $p_{T}$ and $\phi$ are the transverse momentum and
azimuthal angle of each hadron, respectively, which is selected
from pseudorapidity $|\eta|>$ 1 in the final state to avoid
autocorrelation, and the superscript $p$ denotes final momentum
space.

For $v_n^{p}$ determined by the final momentum phase space,
we can {\it even-by-event} correct $v_n^{p}$ into $v_n^{r}$ by
\begin{equation}
v_n^{r} = \left\langle \frac{v_n^{p,(e)}-s_n^{r,(e)}sin[n(\psi_n^{p,(e)} - \psi_n^{r,(e)})]}{R_n^{(e)}} \right\rangle,
\label{eq7}
\end{equation}
where the superscript  $(e)$ denotes "event-wise", $R_n^{(e)} = cos[n(\psi_n^{p,(e)} - \psi_n^{r,(e)})]$ is event-wise event plane resolution,  and $s_n^{r,(e)}=sin[n(\phi-\psi_n^{r,(e)})]$ is event-wise sin-term harmonic coefficient. We found that the contribution from $sin$-term is only approximately 10\% , therefore we neglect the $sin$-term and correct $v_n^{p}$ into $v_n^{r}$ by $v_n^{r} = \left\langle v_n^{p,(e)}/R_n^{(e)} \right\rangle$, which is more operable experimentally.

\begin{figure}[htbp]
\resizebox{8.6cm}{!}{\includegraphics{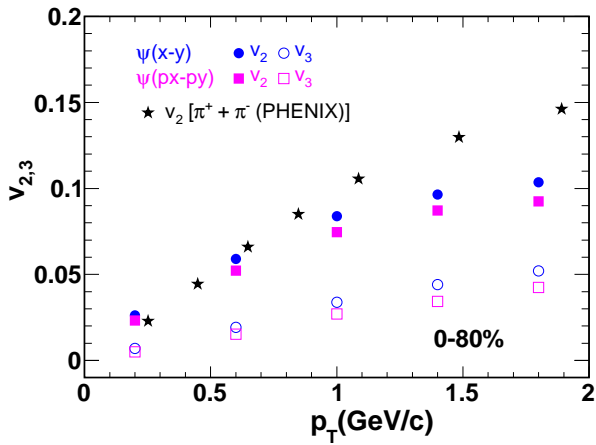}}
\caption{(Color online) Comparison for $v_2$
(solid symbols) and $v_3$ (open symbols) calculated by $\psi_{2,3}^{r}$ defined
in coordinate space (circles) or  $\psi_{2,3}^{p}$ momentum space (squares)
in Au + Au collisions (0-80\% centrality) at $\sqrt{s_{NN}} = 200$ GeV from
the AMPT simulation (3 mb). Note that $v_2$ and $v_3$ defined in
momentum space (squares) have been even-by-event corrected by the event plane
resolution. The PHENIX data are shown by solid
stars~\cite{PHENIXDataPi}.} \label{v2_v3_pt_CE_PE}
\end{figure}

It is essential to check if the  $v_n(p_T)$ calculated by different $\psi_n$
defined in coordinate space and momentum space is similar or not,
because the determination of $\psi_n^{r}$ in coordinate space by
the Eq.(~\ref{eq3}) is not accessible in experiment. The $p_{T}$
dependences of $v_2$ and $v_3$ with respect to $\psi_n$ determined
by initial coordinate and final momentum spaces are shown in
Figure~\ref{v2_v3_pt_CE_PE} together with the PHENIX $v_2$ data
~\cite{PHENIXDataPi}. We observed that $v_2$ and $v_3$ determined
by the final momentum space is very close to the ones with respect to
initial coordinate space. (Note: we check that the differences are
due to $sin-$term contributions in event-by-event resolution
corrections.) Also, the values can basically fit the PHENIX data,
especially at low $p_T$.

Based upon the above observations on  $v_2$ and $v_3$ with
different phase space methods, we conclude that they basically can
present the same results. Therefore in our following calculations,
we apply the initial coordinate space to calculate $\psi_n^{r}$ and
then obtain the corresponding $v_n$.

\subsection{Initial fluctuations and ratio of $v_n/\varepsilon_n$}

\begin{figure}[htbp]
\resizebox{8.6cm}{!}{\includegraphics{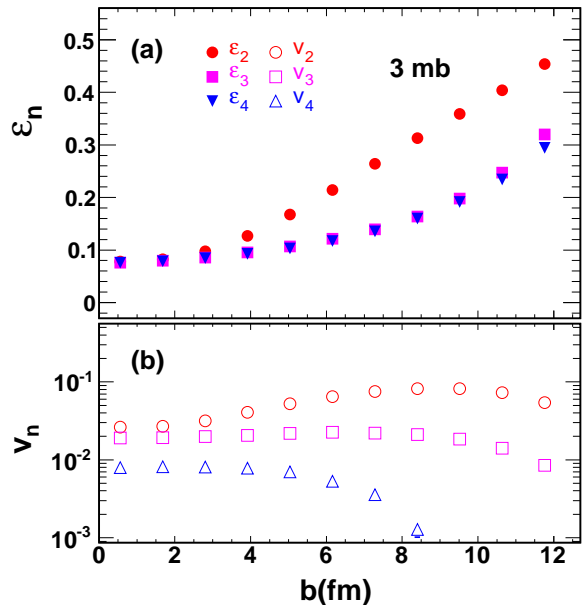}}
\caption{(Color online) $\varepsilon_n$ and $v_n$ in mid-rapidity
as functions of impact parameter for Au + Au collisions at
$\sqrt{s_{NN}} = 200$ GeV from the AMPT simulation with partonic
interaction cross section of 3 mb. } \label{3mb_vn_en_figure}
\end{figure}

The ratio of elliptic flow to eccentricity ($v_2$/$\varepsilon_2$)
has been found to be sensitive to the freeze-out dynamics, the
equation of state (EOS) and viscosity, however,
$v_3$/$\varepsilon_3$ can give more
information~\cite{triangularflow12,ebeqiu,EOS21,hydr3+122,Dipole23,Alverhydro25}.
Figure~\ref{3mb_vn_en_figure} shows the initial $n$-th order eccentricity
$\varepsilon_n$ and final harmonic flow $v_n$ ($n$ = 2, 3 and 4)
in mid-rapidity as a function of impact parameter for Au + Au
collisions at $\sqrt{s_{NN}}$ = 200 GeV from the AMPT model
simulations. The elastic two-body scattering cross section in the
parton cascade process is set to be 3 mb.

\begin{figure}[htbp]
\resizebox{8.6cm}{!}{\includegraphics{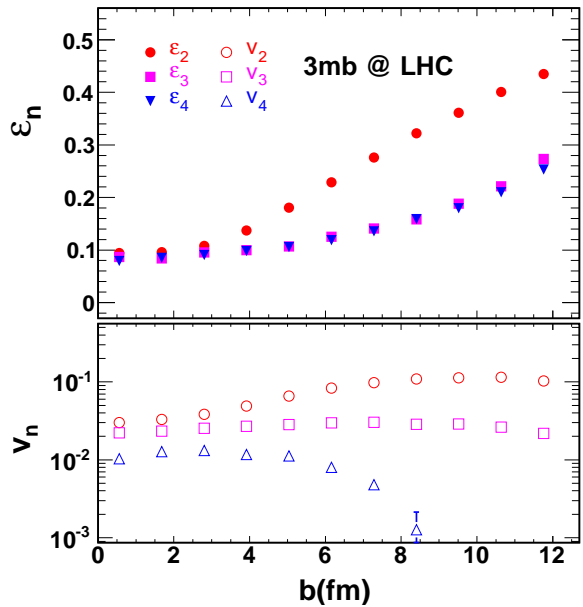}}
\caption{(Color online) Same as Fig.~\ref{3mb_vn_en_figure} but
for Pb + Pb collisions at $\sqrt{s_{NN}} = 2.76$ TeV. }
\label{3mb_vn_en_LCH-figure}
\end{figure}

As presented in Figure~\ref{3mb_vn_en_figure}, the $n$-th order eccentricity
 $\varepsilon_n$ ($n$ = 2, 3 and 4) increases with
impact parameter. $\varepsilon_2$ is larger than $\varepsilon_3$
and $\varepsilon_4$, except in very central collisions where
$\varepsilon_n$ looks similar to each other. It is consistent
with the trend given by Lacey {\it et al.} who applied MC-Glauber model,
but gave a little smaller magnitude for peripheral
collisions~\cite{e2fluc24}. On the other hand, the coefficients of
anisotropic flow $v_n$ ($n$ = 2, 3 and 4) show  rising and falling with impact parameter. Also, $v_n$ has a larger magnitude for
lower harmonic than higher harmonic.

Similarly, Figure~\ref{3mb_vn_en_LCH-figure} shows the initial
geometry deformation $\varepsilon_n$ ($n$ = 2, 3 and 4)  for Pb +
Pb collisions at $\sqrt{s_{NN}} = 2.76$ TeV, which demonstrates
very similar behaviors as the RHIC energy.

\begin{figure}[htbp]
\resizebox{8.6cm}{!}{\includegraphics{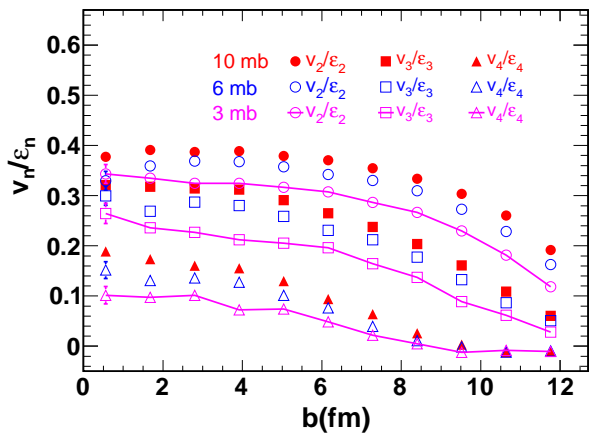}}
\caption{(Color online) $v_n/\varepsilon_n$ (circles for $n$=2, squares for $n$=3, and triangles for $n$=4) as
a function of impact parameter for Au + Au collisions at
$\sqrt{s_{NN}} = 200$ GeV in mid-rapidity for different parton interaction cross
sections (10 mb, 6 mb and 3 mb). } \label{1063mb_vn_en_figure}
\end{figure}

Once we have $v_n$ and $\varepsilon_n$, we can discuss the ratio
$v_n/\varepsilon_n$ ($n$ = 2, 3 and 4). Figure~\ref{1063mb_vn_en_figure} shows impact parameter
and  partonic cross section dependences of $v_n/\varepsilon_n$  for Au + Au collisions at
$\sqrt{s_{NN}} = 200$ GeV. The value of ratios decreases with
impact parameter, which implies that the conversion from the
initial geometry asymmetry to final momentum anisotropy is less
efficient for peripheral collisions than for central collisions.
And for higher harmonics, there is also less conversion efficiency.
The trend for $v_n/\varepsilon_n$ as a function of impact
parameter looks similar for the different partonic interaction
cross sections of 3, 6 and 10 mb. However, the magnitude of
$v_n/\varepsilon_n$ decreases with the cross
section, which reveals that the conversion from the initial geometry
asymmetry to the final momentum anisotropy becomes weaker for a
smaller cross section. This  indicates that frequent
parton-parton collisions help the system to develop the harmonic collectivity.

\begin{figure}[htbp]
\resizebox{8.cm}{!}{\includegraphics{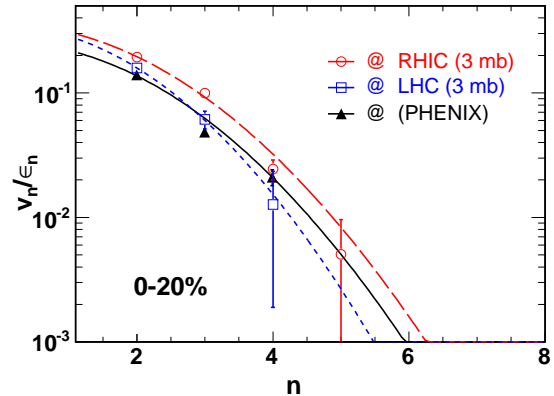}}%
\caption{(Color online) Harmonic order dependence of $v_n$/$\epsilon_n$ for
Au + Au collisions at $\sqrt{s_{NN}} = 200$ GeV (circles) and Pb +
Pb collisions at  $\sqrt{s_{NN}} = 2.76$ TeV (squares) from the
AMPT simulation (3 mb). PHENIX data are shown by triangles. The centrality is 0-20$\%$ and $p_T <$
0.55 GeV/c. The curves are exponential fitting functions.
} \label{vis-harizon}
\end{figure}

From Figure~\ref{1063mb_vn_en_figure}, we also saw that the $v_n/\varepsilon_n$  becomes smaller
for higher harmonic order, this may reflect the viscous damping. Recently, it was claimed that the
relative magnitude of the higher-order harmonics ($v_n$,$n\geq
3$) can provide additional constraints on both the
magnitude of $\eta/s$ and the determination of initial condition
~\cite{Shuryak,e2fluc24,Alverhydro25}.
 Fig.~\ref{vis-harizon} shows the $n$-dependence of
$v_n$/$\epsilon_n$ for Au + Au collisions at $\sqrt{s_{NN}} = 200$
GeV and Pb + Pb collisions  at  $\sqrt{s_{NN}} = 2.76$ TeV for
0-20 $\%$ centrality and low $p_T$ region ($p_T<$ 0.55 GeV/c) (3
mb) with corresponding exponential fitting functions. Compared
with PHENIX data , only the trend can be reproduced.

\subsection{ $p_T$ dependence of $v_n$ with different partonic cross sections and comparisons with the data}

\begin{figure}[htbp]
\resizebox{8.6cm}{!}{\includegraphics{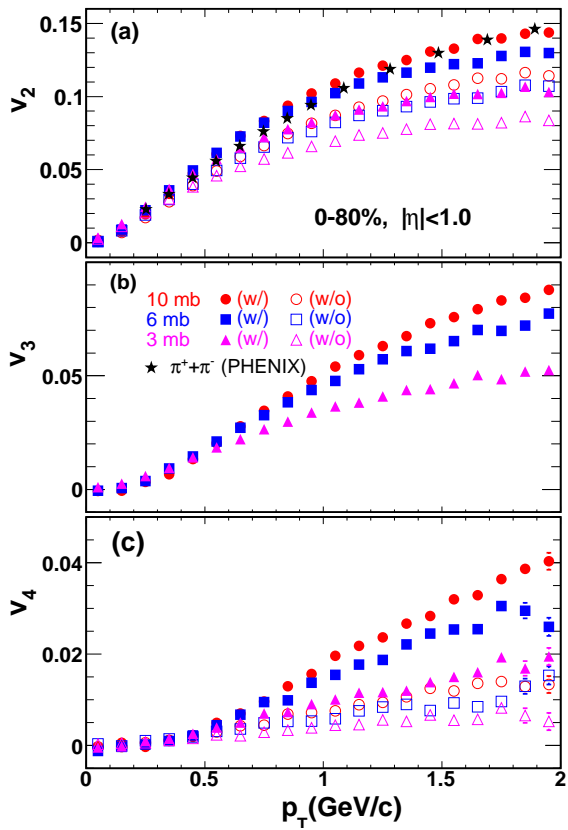}}
\caption{(Color online) (a)-(c): Transverse
momentum dependences of $v_2$, $v_3$,  and $v_4$ for Au + Au
collisions (0-80\% centrality) at $\sqrt{s_{NN}} = 200$ GeV in mid-rapidity
for different cross sections [10 mb (circle), 6 mb (square), and 3
mb(triangle)], where solid symbols are $v_n$ with considering
initial fluctuations and open symbols are those without considering
initial fluctuations. The PHENIX data are shown by solid stars
~\cite{PHENIXPIData}.} \label{v234_pt_data}
\end{figure}

Figure~\ref{v234_pt_data} presents our simulations of $v_2$, $v_3$
and $v_4$ as a function of $p_T$ with different parton interaction
cross sections together with the PHENIX data \cite{PHENIXPIData}.
For triangular flow, it totally arises from the event-by-event
fluctuations of the initial collision geometry, because it
persists zero if without considering the fluctuations. $v_n$ ($n$
= 2, 3 and 4) decreases when parton-parton cross section
decreases. Experimental data of $v_2$ can be described by the large cross
sections (from 3 mb to 10 mb), after one considers of the initial
fluctuations. However, the AMPT model underestimates the data if
without taking the initial fluctuations into account. Recently, Xu
and Ko adjusted more parameters in the AMPT model, which include not only parton
interaction cross section but also the parametrization of the Lund
string fragmentation, and found that a smaller cross section of
1.5 mb is good to describe both the charged particle multiplicity
and elliptic flow~\cite{JunV3}.  In our work, we will not focus on
how to further improve parameters, but we do find that initial
geometry fluctuations significantly affect harmonic flows and
should not be ignored.

\begin{figure}[htbp]
\resizebox{8.6cm}{!}{\includegraphics{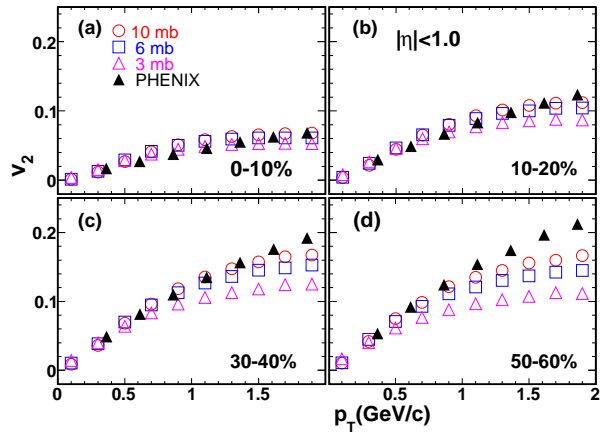}}
\caption{(Color online) $v_2$ as a function of
$p_T$ in Au + Au collisions at $\sqrt{s_{NN}} = 200$ GeV for
different centrality bins (0-10\%, 10-20\%, 30-40\% and 50-60\%)
with different cross sections. Circles, squares and triangles
represent the calculation with 10 mb, 6 mb and 3 mb, respectively. The PHENIX
data shown by solid triangle~\cite{PHENIXData}.}
\label{v2_pt_4cen}
\end{figure}

\begin{figure}[htbp]
\resizebox{8.6cm}{!}{\includegraphics{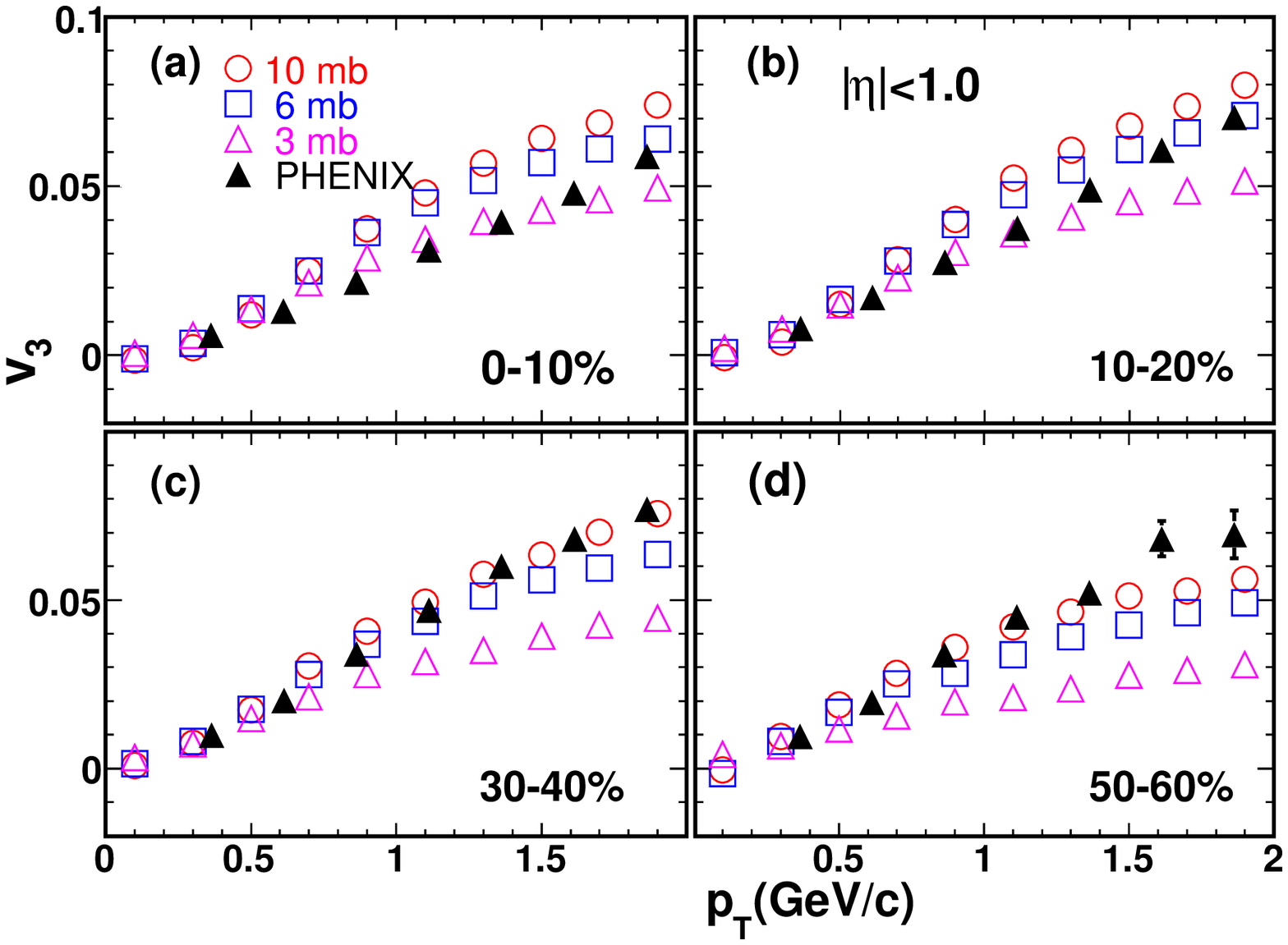}}
\caption{(Color online) Same as Fig.~\ref{v2_pt_4cen} but for $v_3$. }
\label{v3_pt_4cen}
\end{figure}

The transverse momentum dependences of $v_2$ and $v_3$ with
different cross sections in four different centrality bins are
shown in Figure~\ref{v2_pt_4cen} and~\ref{v3_pt_4cen}. The PHENIX data is also accompanied ~\cite{PHENIXData}. For each centrality bin, $v_2$ and $v_3$ increase with the cross section.
For elliptic flow (Fig.~\ref{v2_pt_4cen}), data seem to prefer a
bigger cross section in higher transverse momentum range. In the
case of triangular flow similar trend is present in Fig.~\ref{v3_pt_4cen}, though $v_3$
shows a less centrality dependence than $v_2$, which is consistent
with the trends shown in Figure~\ref{3mb_vn_en_figure}.

\begin{figure}[htbp]
\resizebox{8.6cm}{!}{\includegraphics{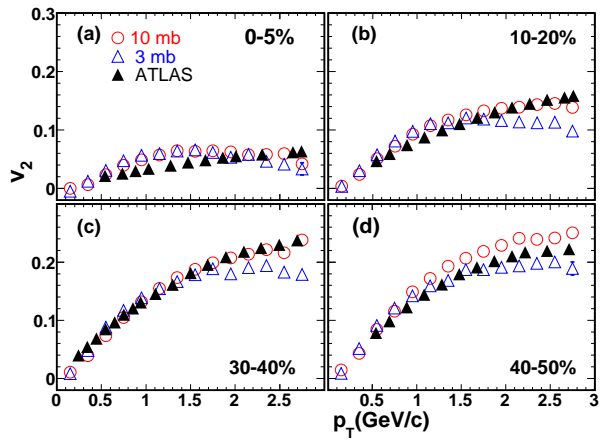}}
\caption{(Color online) $v_2$ as a function of
$p_T$ in Pb + Pb collisions at $\sqrt{s_{NN}} = $ 2.76 TeV from
the AMPT simulations (3 mb and 10 mb) for different centrality
bins (0-5\%, 10-20\%, 30-40\%, and 40-50\%) at midrapidity. Circles and triangles
represent the calculations with 10 mb and 3 mb, respectively.  The ATLAS data shown by solid triangles~\cite{JYJiaLHC}.}
\label{v2_pt_4cenLHC}
\end{figure}

\begin{figure}[htbp]
\resizebox{8.6cm}{!}{\includegraphics{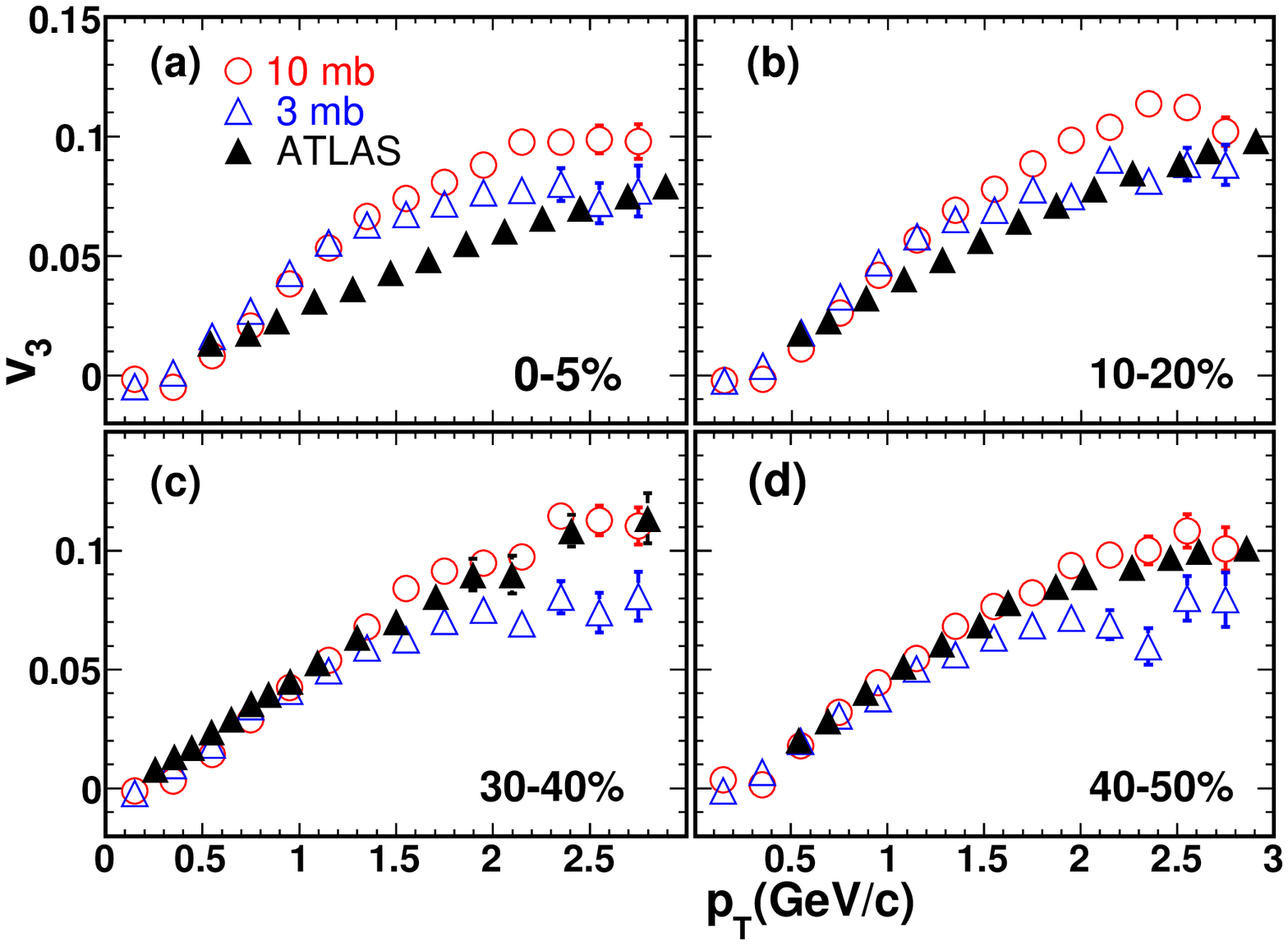}}
\caption{(Color online) Same as Fig.~\ref{v2_pt_4cenLHC} but for $v_3$.}
\label{v3_pt_4cenLHC}
\end{figure}

The transverse momentum dependences  of $v_2$ and $v_3$ are also
calculated for four different centrality bins in Pb + Pb
collisions at $\sqrt{s_{NN}} = $ 2.76 TeV for LHC energy, which
are shown in Figure~\ref{v2_pt_4cenLHC} and~\ref{v3_pt_4cenLHC} together with the ATLAS data \cite{JYJiaLHC}. Similarly, $v_2$ and $v_3$ increase with the cross section from 3
mb to 10 mb, which can basically describe the ATLAS data.

Even though a general behavior of $p_T$-dependent $v_n$ can be
nicely demonstrated by the comparison of our calculations with the
data, we found that the AMPT simulations can only describe the trend of
the data qualitatively. Actually, AMPT can not describe the $v_2$
and $v_3$ data simultaneously with a same cross section. For
example, the 3mb results describe the $v_2$ and $v_3$ for 0-10\% centrality
but underpredict other centralities at RHIC energy. The 10mb
data describe the $v_2$ for 10-20\% centrality but are unable to
reproduce $v_3$ for the same centrality. Also, the 10mb results
describe the $v_2$ and $v_3$ for 30-40\%, but underpredict the high $p_T$
data for 50-60\% centrality. For LHC
data, no AMPT calculations can describe the $v_2$ and $v_3$ for the
0-10\% centrality range.

\subsection{NCQ-scaling of $v_n$ at RHIC energy}

For elliptic flow, a mass ordering (the heavier the hadron mass,
the smaller the $v_2$) and a NCQ-scaling (baryon versus meson)
have been observed at low and intermediate $p_T$, respectively, in
Au + Au collisions at $\sqrt{s_{NN}}$ = 200
GeV~\cite{STARmassord260}. The observed NCQ-scaling ($v_2$/$n_q$
vs $KE_T$/$n_q$) reveals a universal scaling of $v_2$ for all
identified particles over the full transverse kinetic energy
($KE_T$) range, which is more pronounced rather than
$p_T$~\cite{strangeBaryonmassorder5,STARMassOrder6,PHENIXNCQ7,PHENIXphiNCQ8,
tianjianNCQ9,STAR08v2NCQ26,Jinhui4}. Such scaling indicates
that the collective elliptic flow has been developed during the
partonic stage and the effective constituent quark degree of
freedom plays an important role in hadronization process. For
higher even-order harmonics, $v_4$ and $v_6$ etc, appear to be scaled as
$v_n$ $\propto$ $v_2^{n/2}$ ~\cite{STARData} and their NCQ-scaling
has also been suggested in Ref.~\cite{VoloshinvnNCQ29}.
Even in very low energy heavy ion collisions, the $v_2$-scaling
and the $v_4/v_2^2$-scaling have been suggested
for light nuclear clusters in nucleonic level interaction \cite{Yan}. Instead
scaling by the number of constituent quarks ($n_q$) for $v_2$, the
measured data $v_4$, however,  seems to be scaled by
$n_q^{2}$~\cite{v4NCQ160}. It is interesting to check if these scaling relations
are still valid for the $v_n$ calculations in which the initial fluctuations are
taken into account, including the odd harmonics, such as $v_3$.

\begin{figure}[htbp]
\resizebox{8.6cm}{!}{\includegraphics{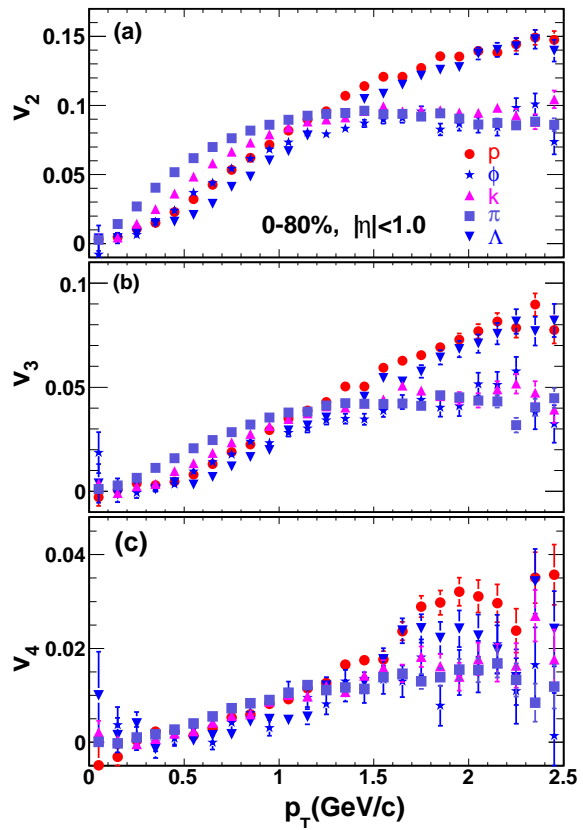}}
\caption{(Color online) (a)-(c): Transverse momentum dependences
of $v_2$, $v_3$, and $v_4$ for different hadron species in Au + Au
collisions (0-80\% centrality) at $\sqrt{s_{NN}} = 200$ GeV in mid-rapidity
from the AMPT simulations (3 mb), with considering of initial
fluctuations. } \label{massorder_3mb}
\end{figure}

\begin{figure}[htbp]
\resizebox{8.6cm}{!}{\includegraphics{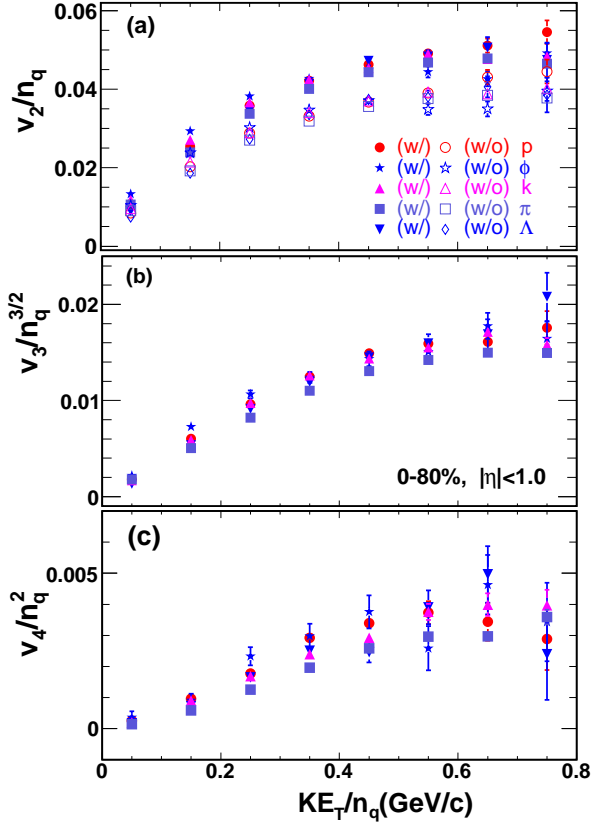}}
\caption{(Color online) (a)-(c): $v_2/n_q$, $v_3/n_q^{3/2}$ and
$v_4/n_q^{2}$ as a function of $KE_T/n_q$ in Au + Au collisions
(0-80\% centrality) at $\sqrt{s_{NN}} = 200$ GeV from the AMPT simulations (3
mb), where the solid symbols are the results for considering initial
fluctuations (w/), while the open ones are for without considering
initial fluctuations (w/o).} \label{NCQ_3mb}
\end{figure}

Figure~\ref{massorder_3mb} presents $v_2$, $v_3$ and $v_4$ of
different types of hadrons in mid-rapidity for Au + Au collisions
(0-80\% centrality) at $\sqrt{s_{NN}} = 200$ GeV. Figure~\ref{massorder_3mb}
(a) shows that $v_2$ preserves an obvious mass ordering in
relatively low $p_T$ region, and hadron type grouping in
intermediate $p_T$ region, even after considering event-by-event
fluctuations. Similarly, $v_3$ and $v_4$,
[Figure~\ref{massorder_3mb} (b) and (c)] also present a mass
ordering in the low $p_T$ region. The study on $v_n$ of different hadron species will
give more information about the initial geometry and the viscosity
of hot and dense matter~\cite{Alverhydro25}.

As shown in Figure~\ref{NCQ_3mb}(a), $v_2$ scaled by the number of
constituent quarks ($v_2/n_q$) as a function of the transverse
kinetic energy ($KE_T = \sqrt{p_T^2 + m^2} - m$) scaled by the
number of constituent quarks ($KE_T/n_q$) shows a universal
scaling regardless of the initial geometry fluctuations are taken into
consideration or not. The only difference is that the initial
fluctuations enhance the value of $v_2/n_q$. Therefore, the
initial fluctuations have little effect on the  breaking of  the NCQ-scaling
for elliptic flow. Figure~\ref{NCQ_3mb} (b) and (c) display
$v_3/n_q^{3/2}$ and $v_4/n_q^{2}$ for all hadrons as  a function of
$KE_T/n_q$, respectively, when the initial fluctuations are
considered. Form the above results, it seems that $v_n$ can still
be roughly scaled by $n_q^{n/2}$ for all hadrons as a function of
$KE_T/n_q$. Of course, the scaling behavior is not perfect within the present statistics.
For example,  the amount of spread between different particle species
is less than 10\% for the $v_2$-scaling, it is less than 20\% for the $v_3$-scaling, but it can
reach 20-30\% for the $v_4$-scaling.

In order to understand possible origin of the NCQ-scaling of $v_n$
for different mesons and baryons, we also check  $v_n$ of $u$, $d$
and $s$-quarks as a function of $p_T$ or $KE_T$. As expected,
there exists similar NCQ-scaling of  $v_n$ ($n$=2-4) for all those
constituent quarks. Furthermore, we  find that the values of
$v_n/{n_q}^{n/2}$ of different hadrons are similar to the
values of $v_n$ of $u, d, s$-quarks,
which reflects that the NCQ-scaling of $v_n$ for different hadrons
stems from partonic level.

\begin{figure}[htbp]
\resizebox{8.6cm}{!}{\includegraphics{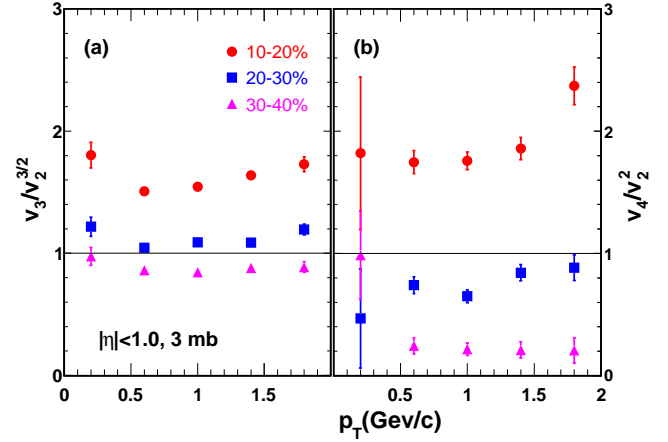}}
\caption{(Color online) The ratios of $v_3/v_2^{3/2}$ (a) and
$v_4/v_2^2$ (b)  as a function of $p_T$ for three different
centrality bins (10-20\%, 20-30\%, and 30-40\%) in Au + Au
collisions at $\sqrt{s_{NN}} = 200$ GeV from the AMPT simulations
(3 mb).} \label{v234Ratio_ptcen}
\end{figure}

Furthermore, the ratios of $v_3/v_2^{3/2}$  and $v_4/v_2^2$ as
functions of $p_T$ for three different centrality bins (10-20\%,
20-30\%, and 30-40\%) in Au + Au collisions at
$\sqrt{s_{NN}} = 200$ GeV are shown in
Figure~\ref{v234Ratio_ptcen}. It shows that the both ratios of
$v_3/v_2^{3/2}$ [Figure~\ref{v234Ratio_ptcen} (a)] and $v_4/v_2^2$
[Figure~\ref{v234Ratio_ptcen} (b)]  exhibit centrality dependences,
i.e. more central collisions result in more larger ratios.
However, it is almost independent of $p_T$ for each centrality
bin, which is consistent with the scaling of $v_n/n_q^{n/2}$
observed in Figure~\ref{NCQ_3mb}. However, it is difficult to
obtain the relationship between $v_{n,q}$ and $v_{2,q}$, such as
between $v_3$ and $v_2$, in terms of a simple coalescence model in
Ref.~\cite{Coalescence}, because $\psi_3$ is purely determined by
initial geometry fluctuations which is independent of $\psi_2$. It
is interesting that  recently  Lacey {\it et al.} linked such a scaling to
the acoustic nature of anisotropic flow to constrain initial
conditions, $\eta/s$ and viscous horizon~\cite{NCQ_R}. It further
gives more insights on the dynamics of strongly-interacting
partonic matter and constituent quark degree of freedom in
hadronization process.

\subsection{NCQ-scaling of $v_n$ at LHC energy}

\begin{figure}[htbp]
\resizebox{8.6cm}{!}{\includegraphics{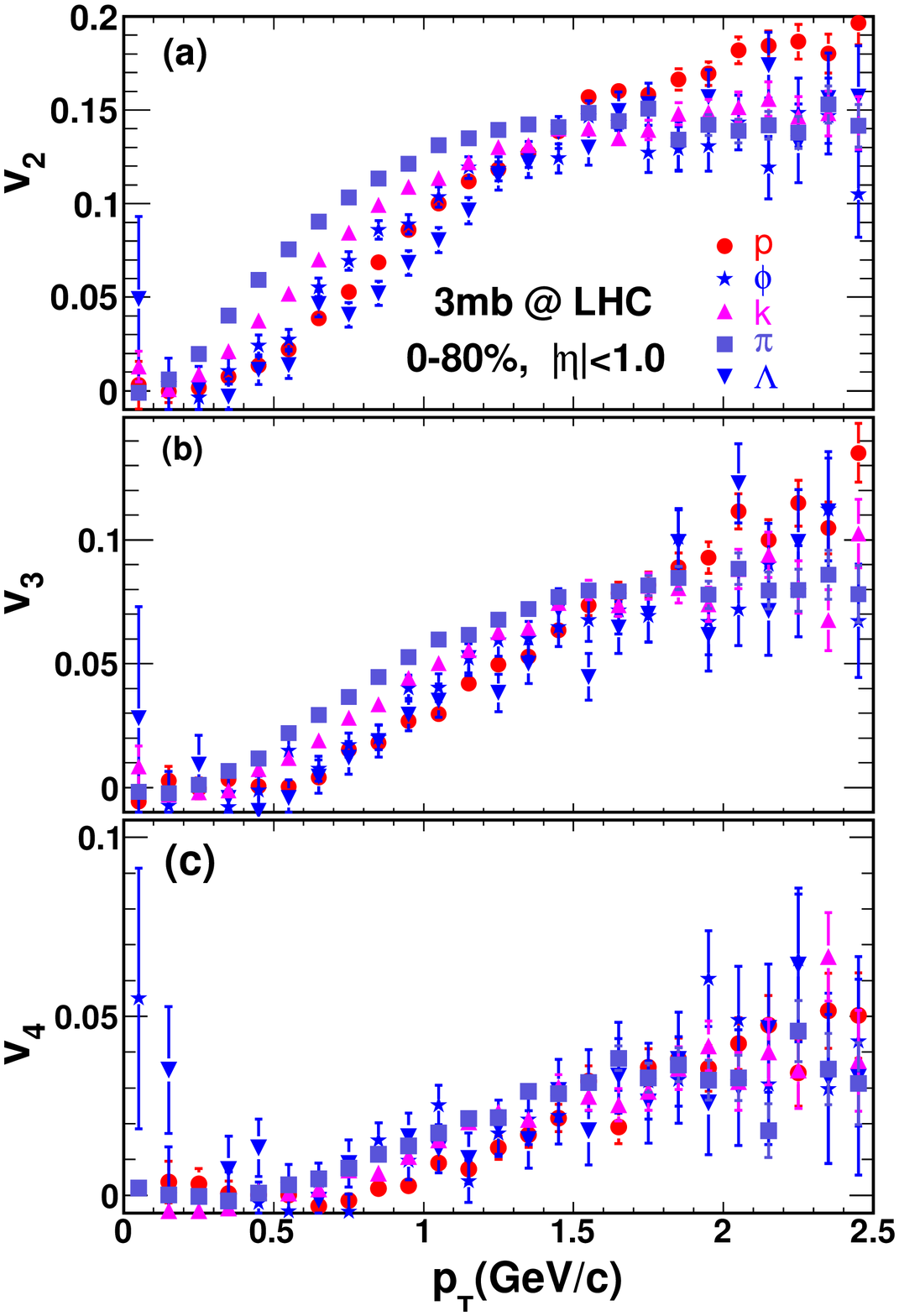}}
\caption{(Color online) (a)-(c): The transverse momentum
dependences  of $v_2$, $v_3$, and $v_4$ for different hadron
species in Pb + Pb collisions (0-80\% centrality) at $\sqrt{s_{NN}} = 2.76$
TeV in mid-rapidity from the AMPT simulations (3 mb), with
considering initial fluctuations. } \label{massorder_3mb_LHC}
\end{figure}

\begin{figure}[htbp]
\resizebox{8.6cm}{!}{\includegraphics{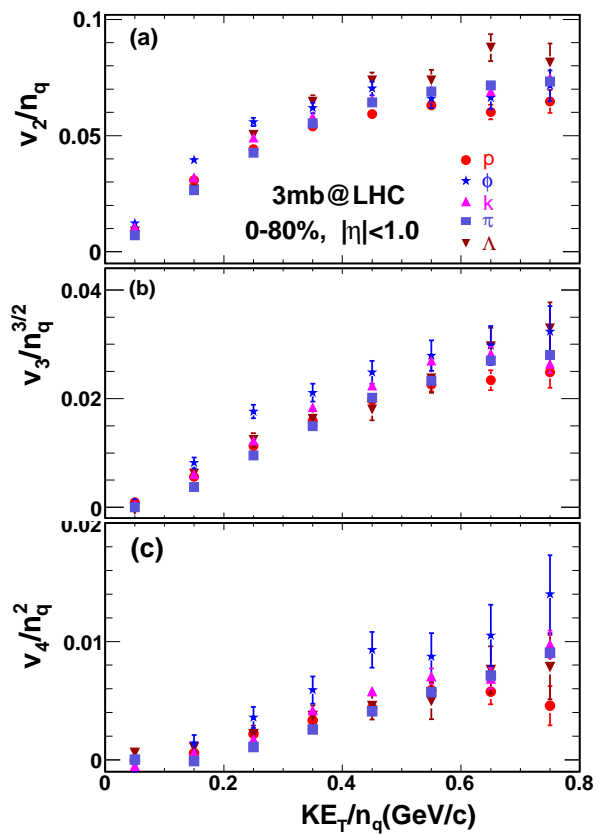}}
\caption{(Color online) (a)-(c): $v_2/n_q$,
$v_3/n_q^{3/2}$ and $v_4/n_q^{2}$ as a function of $KE_T/n_q$ in
Pb + Pb collisions (0-80\% centrality) at $\sqrt{s_{NN}} = 2.76$ TeV from the
AMPT simulations (3 mb).} \label{LHC-NCQ}
\end{figure}

At the same time, the mass ordering in the low $p_T$ region and
the NCQ-scaling in intermediate $p_T$ region of $v_{n}$ are also
investigated at LHC energy in AMPT simulations.
Fig.~\ref{massorder_3mb_LHC} presents the results of $v_2$, $v_3$,
and $v_4$ for different hadron species in Pb + Pb collisions
(0-80\% centrality) at $\sqrt{s_{NN}} = 2.76$ TeV  in mid-rapidity
from the AMPT calculations (3 mb), with considering initial
fluctuations. It  displays that the mass ordering is satisfied,
i.e. $v_n$ decreases from $\pi$, $k$, $p$, $\phi$ to $\Lambda$ in
the lower $p_T$ region (note that $\phi$ is very close to $p$ in
the figure. However, the strict mass-ordering needs $\phi$'s $v_n$
is little less than $p$'s $v_n$). The baryon-meson typing is also
evident above $p_T \sim$ 1.2 GeV. By transformation of $p_T$ to
$KE_T/n_q$ as well as $v_n$ to $v_n/n_q^{n/2}$, the results of
$v_2/n_q$, $v_3/n_q^{3/2}$ and $v_4/n_q^{2}$ as a function of
$KE_T/n_q$  are shown in Fig.~\ref{LHC-NCQ}. Again, the
NCQ-scaling of $v_n$ is roughly kept except for $\phi$ meson whose
$v_n$ is a little larger. Of course, the amount of spread between
different particle species for $v_n$-scaling keeps similar  as
RHIC energy. Comparing with the above $v_n$ results at RHIC, LHC
results are very similar but they reveal larger $v_n$ values than
RHIC's due to stronger partonic interactions at higher energy. But
in general, the partonic matter formed at LHC energy is very
similar to that created at RHIC energy.

As mentioned above,  $\phi$-meson shows a little larger value of
$v_n/{n_q}^{n/2}$  as compared to those of other hadrons in
Fig.~\ref{LHC-NCQ}. Keep in mind that $\phi$-meson is always a
very interesting hadron in previous studies since its mass is
close to proton but it is a multi-strange meson
\cite{Ma-phi,Jinhui4,Chen-PRL,Beganda,Chen-phialign,MaGL-phi}.
This could be understood from the parton's $v_n$ in the same
condition: $v_n$ of $s$-quark displays a slight deviation from the
$u$ ($d$)-quarks (not shown here). The reason could be that the
$v_n$ of heavier strange quarks has a smaller value at low $p_T$
but a larger value at high $p_T$, i.e. the mass ordering of
partonic flow. However, a larger collective
radial flow at LHC energy could push heavier $s$-quark to have
stronger $v_n$. The effect is of course more distinct at LHC
energy because of larger initial partonic pressure. In contrast,
in low energy RHIC run, such as 11.5 GeV/c Au + Au collision,
$s$-quark may not reach full thermalization and therefore result
in a less $v_2$ of  $s$-quark as compared to $u$($d$)-quarks,
which can lead to a smaller $v_2$ of $\phi$, i.e. the violation of
the $v_2$-scaling for the $\phi$-mesons relative to other hadrons
as observed in the STAR data \cite{Beganda} as well as in a simulation \cite{tianjianNCQ9}. Considering that the
$\phi$-meson is coalesced by $s$$\overline{s}$ in the present AMPT
model calculation, it will certainly induce a larger
$v_n/{n_q}^{n/2}$ for $\phi$ in comparison with other hadrons, as
shown in Fig.~\ref{LHC-NCQ}. Unfortunately, the data of $\phi$'s
$v_n$ is not available yet at LHC energy, which is worth waiting
for checking.

Before closing the discussions on the NCQ-scaling of $v_n$ in this
subsection, we remind that the hadronic rescattering process is
not yet taken into account in our calculation. Recently, the ALICE
data shows that proton's $v_2$ and $v_3$ seem to deviate from the
NCQ-scaling of $v_n$ of charged $\pi$ and $K$ \cite{Krzewicki}.
The reason could be stronger final-state interaction for protons.
Detailed model investigations are underway.

\section{Summary}

Within the framework of a multi-phase transport model, we
investigated the different orders of harmonic flows, namely
elliptic flow, triangular flow and quadrangular flow for Au + Au
collisions at $\sqrt{s_{NN}} = 200$ GeV as well as Pb + Pb collisions at $\sqrt{s_{NN}} = 2.76$ TeV
when the initial geometry fluctuations are taken into account.
Basically, the harmonic flow is converted from initial geometry
shape via parton cascade process, and its conversion
efficiency ($v_n/\varepsilon_n$) decreases with the increasing of
harmonic order as well as the decreasing of the partonic cross
section at  both RHIC and LHC energies. Dependences of transverse
momentum, centrality and partonic cross section of the $v_n$ (n=2,
3 and 4) have been studied and compared with data. For each
centrality bin, $v_2$ and $v_3$ increases with cross section,
especially at higher transverse momentum.

Triangular and quadrangular flows also roughly present  a mass ordering in
 low $p_T$ region and the number of constitute quark scaling
in intermediate $p_T$ region, similar to the behaviors of elliptic
flow. Form our results, a NCQ-scaling of $v_n/n_q^{n/2}$ versus
$KE_T/n_q$ for different hadrons holds for harmonic flow ($v_n$,
$n$ = 2, 3 and 4), which can be related to $v_n$-scaling in
partonic level. From all above results, it implies that the formed
partonic matter should be very similar for RHIC and LHC energies.

\section*{Acknowledgements}
The authors thank Dr. B. Johnson, and Dr. S. Esumi for providing
data kindly. The authors also thank Dr. M. Luzum, Dr. J. Y.
Ollitrault, Dr. Y. Zhou and Dr. R. Lacey for helpful discussion.
This work was supported in part by  the National Natural Science
Foundation of China under contract Nos. 11035009, 11005140,  11175232, 10979074,
10875159 and the Knowledge
Innovation Project of the Chinese Academy of Sciences under Grant
No. KJCX2-EW-N01, and the Project-sponsored by SRF for ROCS, SEM.


\end{document}